\documentclass{aa}  

\usepackage{graphicx}
\usepackage{tabularx}
\usepackage{txfonts}
\usepackage{dcolumn}
\usepackage{bm}
\usepackage{url}
\usepackage{hyperref}
\usepackage{subcaption}
\usepackage{xcolor}

\begin{document}

   \title{Cosmological measurement of the gravitational constant $G$ using the CMB, the BAO and the BBN}


   \author{B. Lamine
          \inst{1, \thanks{brahim.lamine@irap.omp.eu}}
          \and
          Y. Ozdalkiran
          \inst{1}
          \and
          L. Mirouze
          \inst{1}
          \and
          F. Erdogan
          \inst{1}
          \and
          S. Ilic
          \inst{2}
          \and
          I. Tutusaus
          \inst{1}
          \and
          R. Kou
          \inst{1,3}
          \and
          A. Blanchard
          \inst{1}
          }

   \institute{Institut de Recherche en Astrophysique et Plan\'etologie (IRAP), Universit\'e de Toulouse, CNRS, UPS, CNES, 14 Av. Edouard Belin, 31400 Toulouse, France
         \and
             Université Paris-Saclay, CNRS/IN2P3, IJCLab, 91405 Orsay (France)
          \and Department of Physics \& Astronomy, University of Sussex, Brighton BN1 9QH, UK
             }


 
  \abstract
{Recent cosmological observations have provided numerous new observations with increasing precision that have led to the era of precision cosmology. The exquisite quality of these observations opens new possibilities towards measuring fundamental constants with good precision and at scales which are complementary to the laboratory ones. In particular, the cosmic microwave background (CMB) temperature and polarization spectra contain a wealth quantity of information, well beyond the basic cosmological parameters. In this paper, we update the precision on a cosmological determination of $G$ by using the latest Planck data release (PR4) in combination with the latest baryon acoustic oscillation (BAO) from the Dark Energy Spectroscopic Instrument (DESI) data release 1 and the BBN prior on the primordial Helium fraction. We demonstrate a precision of $1.8\%$, corresponding to a $\sim25\%$ improvement compared to the literature. This is comparable to the level achieved by Cavendish in 1873 using a torsion balance. However, it is a complementary measurement because it has been obtained under wildly different physical environments compared to the laboratory or even the very nearby Universe. Our analysis takes into account the modification of the primordial Helium fraction predicted by Big Bang nucleosynthesis (BBN), induced by a variation of $G$. We also point out the importance of the polarization data in the final precision and in particular discuss the constraints that can be obtained by considering either the low-$\ell$ or the high-$\ell$ part of the spectra. Within the $\Lambda$CDM model, we find $G=(6.75\pm0.12) \times 10^{-11} \mathrm{m}^3\cdot\mathrm{kg}^{-1}\cdot\mathrm{s}^{-2}$. This measurement is compatible with laboratory ones within one standard deviation. Finally, we show that this cosmological measurement of $G$ is robust against several assumptions made on the cosmological model, in particular when considering a non-standard dark energy fluid or non-flat models.}


   \maketitle
%

\section{Introduction}

Despite being the oldest fundamental constant of our standard model, the gravitational constant $G$ has not been used to define the kilogram in the new international system of units~\cite{si-brochure}. The Planck constant $h$ (not to be confused with the reduced Hubble constant) has instead been selected and fixed at a specific value. Consequently, the gravitational constant must be determined experimentally, and unfortunately it remains the least precisely measured fundamental constant. This enigma can be readily attributed to the exceptionally feeble nature of the gravitational force when compared to the other fundamental forces in the Universe. Therefore, one rapidly arrives at situations where spurious non-gravitational effects surpass the relevant gravitational effect when trying to measure $G$ with high precision and accuracy. According to the Committee on Data for Science and Technology (CODATA 2018), laboratory experiments enable to measure Newton's gravitational constant $G_{\text{codata}}=6.674\,30(15)\times10^{-11}\mathrm{m}^3\cdot\mathrm{kg}^{-1}\cdot\mathrm{s}^{-2}$ with a relative uncertainty of (only) $2.2\times 10^{-5}$~\citep{codata}. Moreover, these laboratory measurements suffer from not yet completely understood terrestrial effects that affect the final precision of $G$~\citep{fixlerAtomInterferometerMeasurement2007,andersonMeasurementsNewtonGravitational2015,xuePrecisionMeasurementNewtonian2020}, which makes an independent measurement of $G$ at cosmological scales interesting.

To achieve a high-precision measurement of $G$, it is imperative to consider physical experiments where non-gravitational effects are effectively under control, and gravitational forces assume a prominent role, thus permitting the disentanglement of non-gravitational contributions from their gravitational counterparts~\citep{uzanFundamentalConstantsTheir2003,uzanVaryingConstantsGravitation2011}. Interestingly enough, this is a situation that can be met in cosmology during the baryon-photon plasma phase before recombination, where both gravitation effects and Thomson scattering of photons by electrons are present~\citep{PhysRevD.67.063002}. Indeed, the physics involved during this period are supposedly well understood, as illustrated by the very good fit of our theoretical model to the measurements of the cosmic microwave background (CMB) temperature and polarization spectra, as well as all the internal consistency checks that can be performed~\citep{Planck2018}. In particular, it is quite spectacular to realize that the temperature of the CMB itself can be inferred solely from its fluctuation spectrum, as demonstrated in~\cite{ivanovTensionTension2020}. Since the physics of recombination involve both non-gravitational and gravitational physics, they allows for a sensitivity to $G$, which is not degenerate with other parameters such as the energy densities. Hence cosmological measurements of $G$ using the CMB have been performed in the literature, either from (phenomenological) model-independent studies~\citep{PhysRevD.67.063002,umezuCosmologicalConstraintsNewton2005,PhysRevD.75.083521,PhysRevD.80.023508,Bai2015,wangConstraintsNewtonConstant2020b,sakrCanVaryingGravitational2022} or for specific models such as scalar-tensor theories~\citep{riazueloCosmologicalObservationsScalartensor2002,nagataWMAPConstraintsScalartensor2004,ballardiniCosmologicalConstraintsGravitational2022a,bragliaLargerValueEvolving2020a,oobaCosmologicalConstraintsScalartensor2017a}. Even if not competitive with laboratory measurements, a cosmological determination of $G$ is complementary in terms of scales. Indeed, many extended models of general relativity (GR) predict either a (space-)time variation of this constant at cosmological scales~\citep[the reference model for that being the Jordan-Brans-Dicke model,][]{1959ZPhy..157..112J,1961PhRv..124..925B,sola2020bransdicke}, or a different value for different spatial scales~\citep{bertschingerDistinguishingModifiedGravity2008}. Some screening mechanisms allow us to recover the laboratory measurements~\citep{JOYCE20151}. In particular, if the gravitational constant inferred from cosmology turns out to be different from the local one, this would point towards a modification of gravity at large scale. Finally, testing the value of $G$ in extreme conditions such as the ones at the very beginning of the Universe, where the energy content is dominated by radiation, is also extremely interesting. 

This paper presents updated constraints on a cosmological measurement of the gravitational constant $G$. Our focus is specifically on high-redshift probes, and we exclude type-Ia supernovae (SNIa) data from our analysis. This exclusion stems from the ongoing debate surrounding the influence of the gravitational constant on the intrinsic luminosity of SNIa. Indeed, the straightforward expectation is that variations in $G$ would affect the intrinsic luminosity due to its presumed correlation with the Chandrasekhar mass $M_{\text{Ch}},$ itself proportional to $G^{-3/2}$. Consequently, an increase in $G$ would result in a diminished intrinsic luminosity~\citep{gaztanagaBoundsPossibleEvolution2001a,garcia-berroVARIATIONGRAVITATIONALCONSTANT2006a,mouldConstrainingPossibleVariation2014,zhaoConstrainingTimeVariation2018}. However, a study presented in~\citet{wrightTypeIaSupernovae2018}~\citep[confirmed in][]{ruchikaGravitationalConstantTransition2023}
conducted a comprehensive examination of the impact of altering the gravitational constant on SNIa. Utilizing semi-analytical models, they discovered that when accounting for both the aforementioned effect and the standardization procedure, in which the light curve is stretched to match a predefined template, the net effect of increasing the gravitational constant $G$ is to amplify the peak luminosity. This result runs counter to the initial intuitive expectation. Hence, we have chosen to narrow our focus exclusively to high-redshift probes, given the uncertainties associated with SNIa measurements in the context of varying $G$.

We consider the latest PR4 analysis of Planck data~\citep{tristramCosmologicalParametersDerived2024}, which represents an improvement with respect to Planck 2018~\citep{Planck2018}. We will use the term P20 to refer to this PR4 data analysis, while P18 will refer to Planck 2018. Combining with the latest baryon acoustic oscillation (BAO) data from DESI as well as a BBN prior for the primordial Helium fraction $Y_{\text{He}}$ from~\cite{oliveImpactCurrentResults2021}, we demonstrate a $1.8\%$ percent precision in the determination of $G$. This value is compatible with laboratory measurements within one standard deviation, therefore not pointing towards a modification of our gravitational laws at large scales. The addition of the BAO measurements only marginally improves the constraint on $G$, except when considering non-flat $\Lambda$CDM models. Our analysis properly takes into account the modification of the primordial Helium fraction predicted by the Big Bang nucleosynthesis (BBN) due to the variation of $G$. We also point out the importance of the polarization data in the final precision, fostering the need for next-generation polarization measurements such as the ones that will be provided by LiteBIRD~\citep{litebirdcollaborationProbingCosmicInflation2023} or CMB-S4~\citep{abazajianCMBS4ScienceCase2019} experiments from the ground, including the South Pole telescope (SPT)-3G~\citep{bensonSPT3GNextgenerationCosmic2014} and the Simons Observatory~\citep{adeSimonsObservatoryScience2019}. We also discuss the constraints that can be obtained by considering either the low-$\ell$ or the high-$\ell$ part of the spectrum. Finally, we show that this cosmological measurement of $G$ is robust against several assumptions made on the cosmological model, in particular when considering a non-standard dark energy fluid or non-flat models. This is particularly important in the context of the slight internal tension in the Planck 2018 data concerning the flatness of the universe~\citep{handleyCurvatureTensionEvidence2021,divalentinoPlanckEvidenceClosed2020a}, or also with the recent evidence by the combination of CMB and BAO measurements from the Dark Energy Spectroscopic Instrument (DESI) of an evolving dark energy~\citep{desicollaborationDESI2024VI2024a}.

In Sect.~\ref{recombination}, we briefly discuss how the gravitational constant affects the temperature and polarization power spectra of the CMB. The methods and results are presented in Sect.~\ref{results}, where we discuss our measurements of the gravitational constant and show that the cosmological model assumptions have no significant impact on the inferred value of~$G$. We present our conclusions in Sect.~\ref{conclusions}.

\section{Effect of changing the gravitational constant in the CMB anisotropies and the light element abundancies (BBN)}
\label{recombination}

In this section we characterize the effect of changing the gravitational constant $G$ in the CMB spectra, both in temperature and polarization. For convenience and in accordance with the literature, we define a free parameter $\lambda_G$ such that $G = \lambda_G^2 G_{\text{codata}}$. The first Friedmann-Lemaître equation then reads 
\begin{equation}
    H^2 = \frac{8\pi \lambda_G^2 G_{\text{codata}}}{3}\sum_i \rho_i\,,
\label{Friedmann}
\end{equation}
where $\rho_i$ is the energy density of species $i$. This model will be denoted $\lambda_G\Lambda$CDM. The Hubble factor $H$ is clearly proportional to $\lambda_G$, which led many authors to look to a potential solution of the Hubble tension thanks to a modification of $G$~\citep{sakrCanVaryingGravitational2022}. These solutions unfortunately fail due to the exquisite constraints imposed by the CMB, which we are going to discuss below.

\begin{figure}[h!t]
    \centering
    \includegraphics[width=\linewidth]{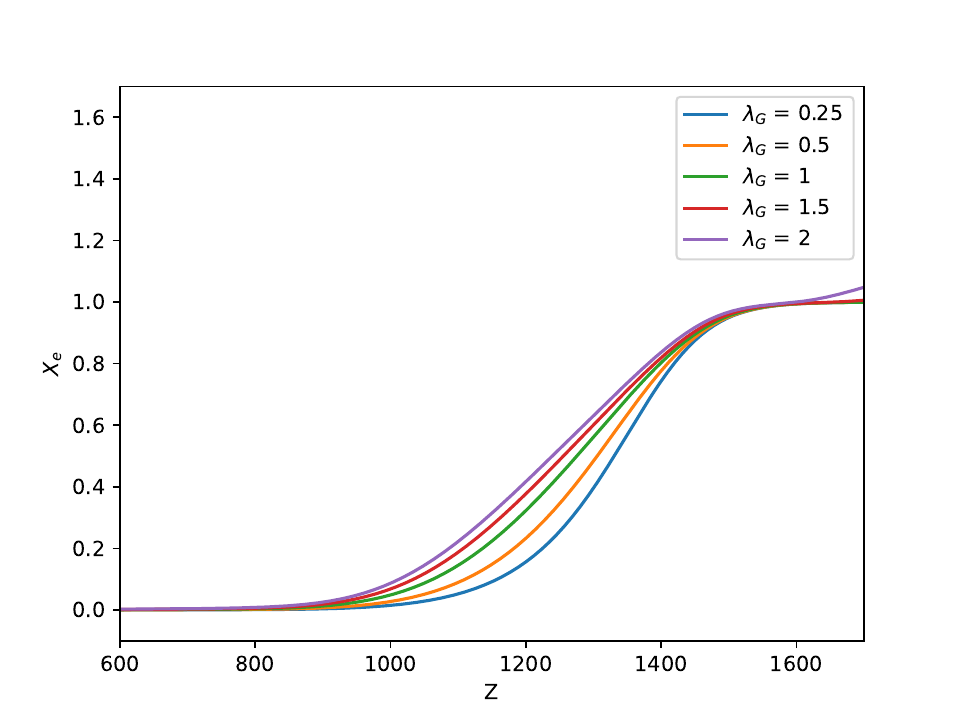}
    \caption{Evolution of the ionization fraction $x_e(z)$ with respect to the redshift, for different values of $\lambda_G$ and fixed physical densities. One sees that increasing $G$ delays the time of recombination.}
    \label{fig:x_e}
\end{figure}

\begin{figure}[h!t]
    \centering
    \includegraphics[width=\linewidth]{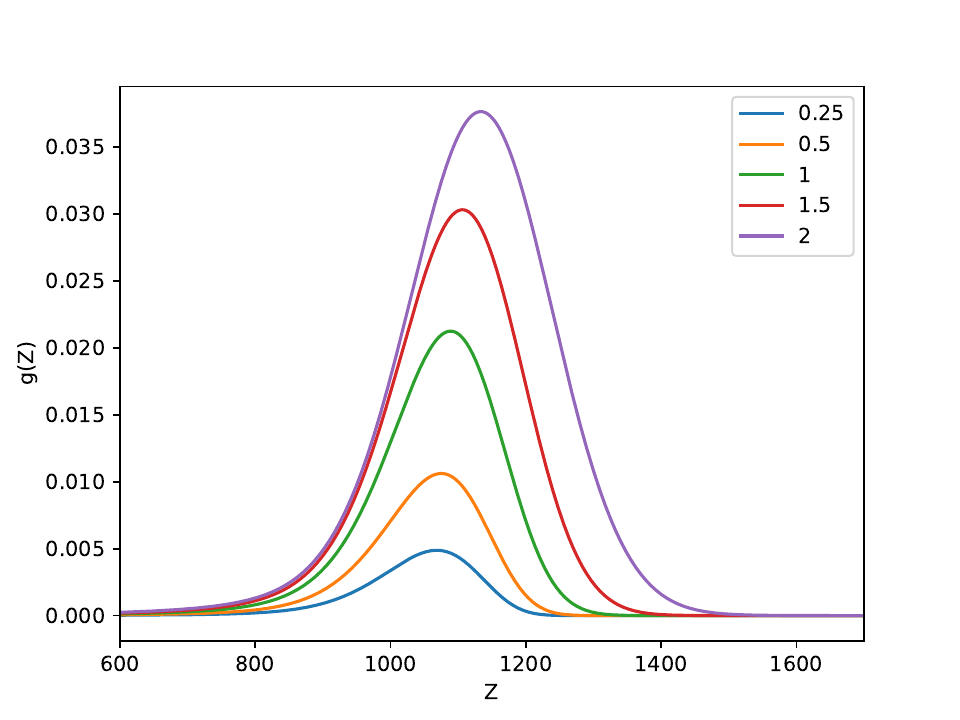}
    \caption{Evolution of the visibility function with respect to the redshift, for different values of $\lambda_G$ and fixed physical densities. The redshift of decoupling (defined as the maximum of the visibility function) increases when $\lambda_G$ increases, so decoupling happens earlier. The width of the decoupling also increases with $\lambda_G$.}
    \label{fig:visibility}
\end{figure}

It has already been pointed out in previous studies that gravity has no preferred scale. Therefore, when considering only gravitational effects, modifying $G$ can be reabsorbed into a rescaling of the wavevectors~\citep{PhysRevD.67.063002}. Since the initial power spectrum of primordial perturbation is assumed to be a power law $P_i(k)=k^{n_{\text{s}}-1}$, a change of $G$ would then be reabsorbed in the amplitude of initial primordial fluctuations and eventually the spectral index $n_{\text{s}}$~\citep{wangConstraintsNewtonConstant2020b}, therefore leaving no net influence of $G$ in the physical observables. Hopefully, non-gravitational physics occur during recombination, in particular Thomson scattering of photons on electrons. The effect of $G$ can then be seen in the interplay between the expansion rate $H\propto \lambda_G$ and the Thomson interaction rate $\dot{\kappa}(z)= an_e\sigma_T$, where $a$ is the scale factor, $n_e$ the free electron number density, and $\sigma_T$ the Thomson cross section. The CMB spectra is sensitive to the ratio $\dot{\kappa}(z)/H(z)$. Therefore, a change of the expansion rate via a change of $G$, without accordingly changing the Thomson rate, will leave an impact on the CMB~\citep{greeneThomsonScatteringOne2023}. More precisely, it has been shown in~\citet{PhysRevD.80.023508} that increasing $G$ delays the time of recombination and increases the recombination duration, as can be seen from the evolution of the ionization fraction in Fig.~\ref{fig:x_e}. That is because the expansion rate would be higher, making it more difficult for electrons to recombine with protons to produce Hydrogen. At the same time, increasing $G$ advances the moment of decoupling by raising the redshift corresponding to the peak of the visibility function, as well as broadening its width (see the Fig.~\ref{fig:visibility} and also~\cite{richWhichFundamentalConstants2015}). This effect on $g(z)$ contrasts with the effect on $x_e(z)$. Note nevertheless that those effects are rather small, since for example multiplying $G$ by a factor 4 only increases the decoupling redshift by $5\%$.

\begin{figure}
  \centering
  \includegraphics[width=\linewidth]{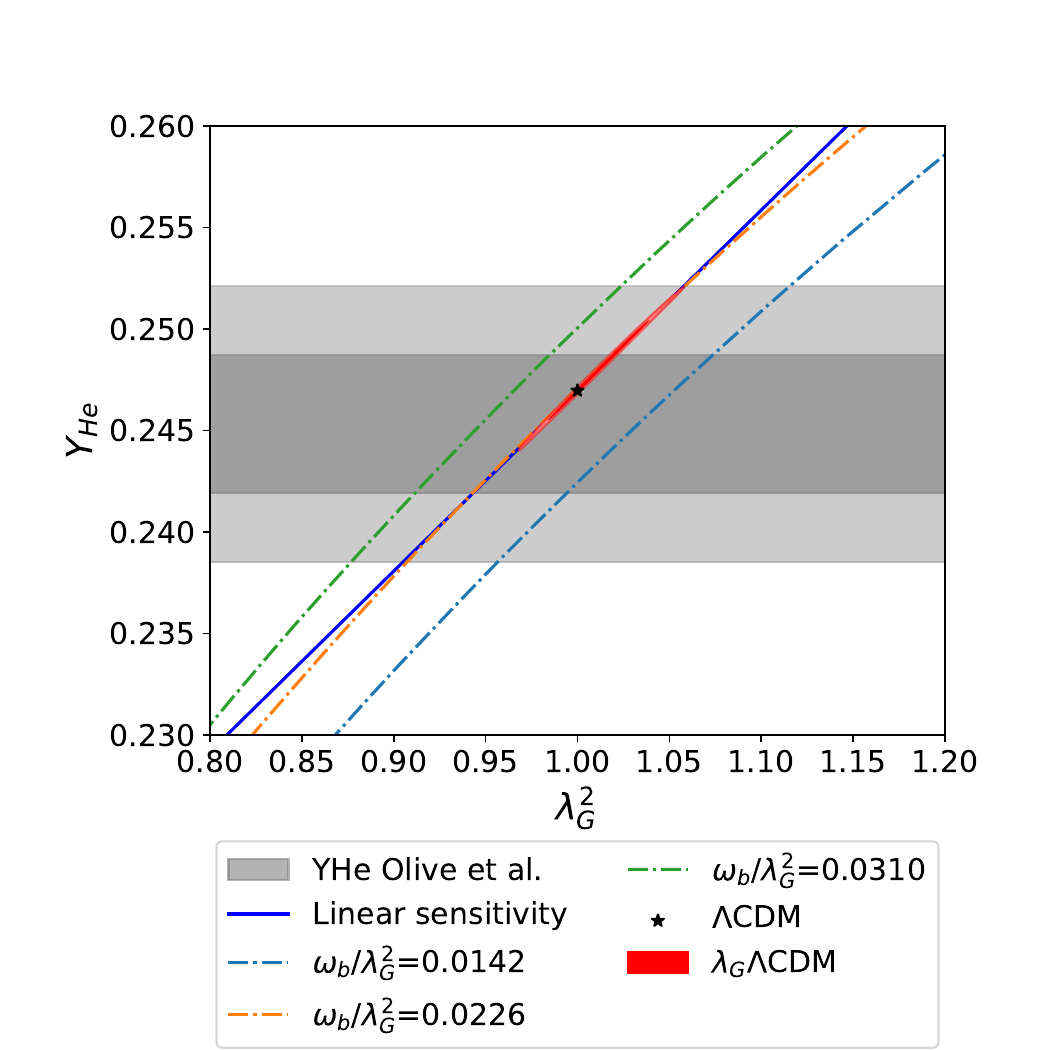}
  \caption{\label{parthenope} Primordial Helium fraction $Y_{\text{He}}$ as a function of $\lambda_G^2$. For each curve, the baryon physical density $\rho_b$, which is proportional to $\omega_{\text{b}}/\lambda_G^2$, has been kept constant while $\lambda_G$ varies. The grey band indicates the observational constraints of~\cite{oliveImpactCurrentResults2021}. The black star is the best-fit standard $\Lambda$CDM model, while the red ellipse corresponds to the contour for the $\lambda_G\Lambda$CDM model fitted with CMB + BAO from DESI + BBN. We observe that this contour is highly degenerate
  in the direction of constant physical density. The blue curve correspond to the linear sensitivity model of~\cite{dentPrimordialNucleosynthesisProbe2007}.
  }
\end{figure}

The same argument holds also for the BBN since the predicted amount of light elements depends crucially on the comparison between the expansion rate $H$ and nuclear reaction rates. Notably, a modification of the expansion rate $H$ changes the time at which various weak and nuclear processes freeze-out. Specifically, the $p + n \longleftrightarrow D + \gamma$ reaction experiences earlier freeze-out if $H$ increases (which could be the case if $\lambda_G>1$, for example). Consequently, it leads to an excessive production of Deuterium and then Helium, when compared to the scenario with $\lambda_G=1$. This is why BBN can independently serve as a valuable tool for estimating the gravitational constant, as discussed in~\cite{accettaNewLimitsVariability1990,copiNewNucleosynthesisConstraint2004a,cyburtNewBBNLimits2005,bambiResponsePrimordialAbundances2005,alveyImprovedBBNConstraints2020a}. As an illustration, Fig.~\ref{parthenope} represents the effect of changing $G$ on the primordial Helium fraction $Y_{\text{He}}$~\citep[similar results as in][]{alveyImprovedBBNConstraints2020a}, for a fixed physical energy density of baryons. In this figure, we represented the linear model of evolution of $G$ from~\cite{dentPrimordialNucleosynthesisProbe2007}, as well as our result stemming from the modified version of the CMB code PArthENoPE~\citep{gariazzoPArthENoPERevolutions2022}. We see that the linear model (labeled linear sensitivity in Fig.~\ref{parthenope}) nicely agrees with our result, as it should. We also remark that the linear approximation of~\cite{dentPrimordialNucleosynthesisProbe2007} is clearly sufficient for our study.

\begin{figure}
  \centering
  \includegraphics[width=\linewidth]{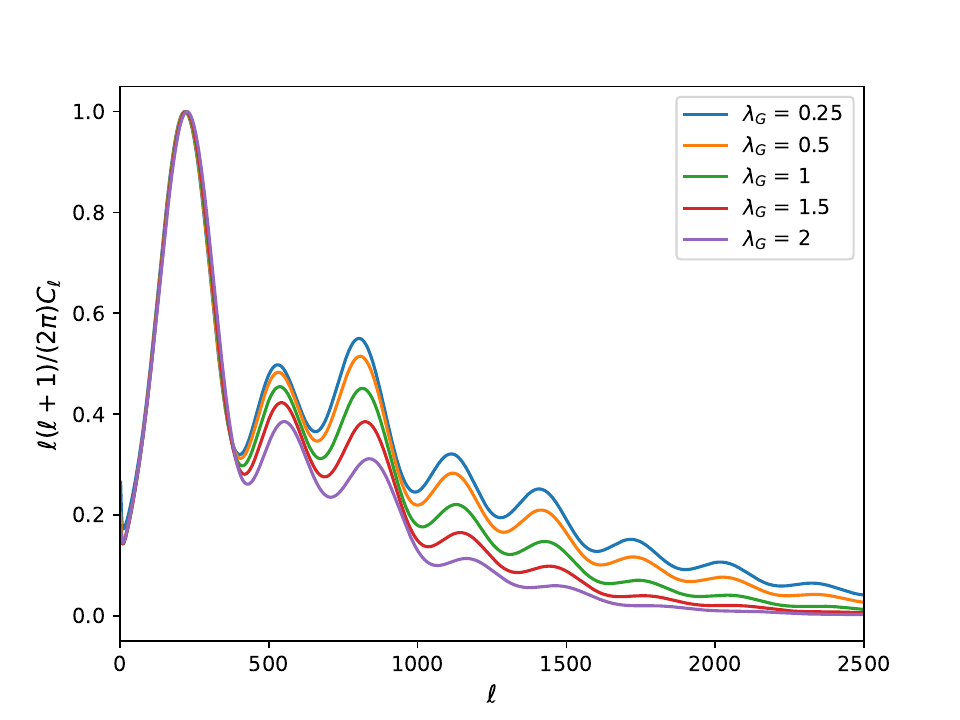}
  \caption{\label{power_spectrum}Effect of $\lambda_G$ on the temperature power spectrum of the CMB. The first peak has been normalized (in the amplitude of the first pic) and the same physical energy densities have been used for each curve. This plot shows the increase of the damping tails with increasing $\lambda_G$.
  }
\end{figure}

Regarding the CMB temperature spectrum, the increase of the recombination duration results in an increase of the damping of the CMB peaks at small scales (characterised by large values of $\ell$). This phenomenon arises due to the propagation of light between hot and cold regions which tends to homogenise the temperature through the effect of viscosity and heat conduction, thereby causing a dampening of the CMB peaks. Small angular scales exhibit a more pronounced impact due to the fact that, during the recombination phase, light is capable of traversing these smaller scales, facilitating heat conduction over them. Conversely, larger scales characterised by wavelengths greater than the distance covered by light throughout the recombination phase do not permit significant heat conduction effects. This effect can readily be seen in the temperature power spectrum of the CMB represented in Fig.~\ref{power_spectrum}. In this plot, the first peaks have been appropriately normalised to facilitate a straightforward comparison of damping behaviours across various values of $\lambda_G$. These curves have been plotted using the same physical energy densities $\rho_i$ (which necessitates a different value for $H_0$ to ensure that $\sum_i \Omega_i = 1$). We clearly observe that higher values of $\lambda_G$ indeed lead to an intensified damping of the peaks at small scales (high $\ell$). This phenomenon forms the basis for extracting the gravitational constant $G$ from the CMB temperature power spectrum. More precisely, one may extract the damping scale $k_D$ from the temperature spectrum, and subsequently compare it to its anticipated value. This expected value is in fact the geometric mean of the horizon size and the mean free path~\citep{huPhysicsMicrowaveBackground1997}, both of which can also be independently extracted from the CMB peaks. The consistency between the observed and expected relationship not only serves as a robust consistency check for CMB data~\citep{huPhysicsMicrowaveBackground1997} but also contributes significantly to the accuracy of the gravitational constant $G$ measurement and its accordance with the standard CODATA value $G_{\text{codata}}$.

\begin{figure}
  \centering
  \includegraphics[width=\linewidth]{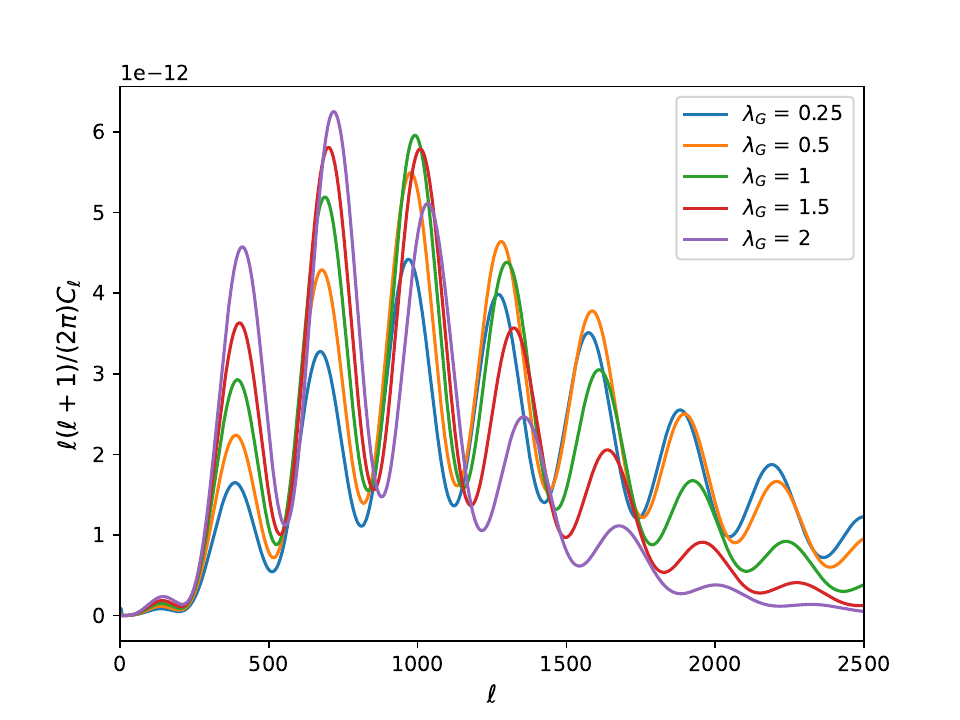}
  \caption{\label{power_spectrum_ee}Effect of $\lambda_G$ on the EE polarization power spectrum of the CMB. The same physical energy densities have been used for each curve. When $\lambda_G$ increases, large $\ell$ are more damped while small $\ell$ are less damped than the $\lambda_G=1$ case.}
\end{figure}

Nonetheless, the impact observed in the temperature spectrum alone does not yield a highly precise measurement of the gravitational constant $G$, as it is evident in Table~\ref{tab:lambdaG} and Fig.~\ref{fig:lambda_G_CMB}, which present constraints derived from the temperature anisotropy (TT) component of the spectrum (see a detailed discussion in the results section). Fortunately, the effect of changing the value of $G$ on the polarization power spectrum (EE) shows an interesting feature that physically justifies the usefulness of combining the polarization data with the temperature one to significantly enhance the precision on the measurement of $G$ (see Table~\ref{tab:lambdaG}). Specifically, apart from the previously-described heightened damping of quadrupolar anisotropies at small scales as $\lambda_G$ increases, there is a concurrent global amplification of quadrupole anisotropies. This occurs because light propagates over longer distances (since the recombination duration increases with $\lambda_G$), resulting in higher level of polarization~\citep{zaldarriagaAnalyticApproachPolarization1995}. The combination of the two previous effects finally leads to an increase of the polarization spectrum for small $\ell$, and a decrease for large $\ell$, as illustrated in Fig.~\ref{power_spectrum_ee}. This behavior has been previously documented and expounded upon in~\cite{PhysRevD.67.063002}. The characteristic scale $\ell^*$ separating these two regimes is approximately $\ell^*\simeq800$. Note than an equivalent feature exists in the TE spectrum (though not represented here). To investigate this further, we conducted separate studies, one focusing solely on the low-$\ell$ portion of the CMB spectrum ($\ell<\ell^*$) and another centered on the high-$\ell$ portion ($\ell>\ell^*$), as elaborated upon in the following section. For this $\ell-$dependency constraints, we used Planck 2018 data instead of the latest PR4 data (since the decomposition in low $\ell$ and high $\ell$ is not yet available in the PR4 likelihood).

\section{Results}
\label{results}

We used a modified version of the public Boltzman code CLASS~\citep{blasCosmicLinearAnisotropy2011} to generate temperature and polarization power spectra that depend on the value of $G$. The code is publicly available~\footnote{\url{https://github.com/yacobozdalkiran/CLASS_mod}}. Additionally, we took into account the modification of the primordial Helium fraction $Y_{\text{He}}$ produced during the BBN phase. This modification was achieved by adapting the PArthENoPE code~\citep{gariazzoPArthENoPERevolutions2022}. Subsequently, we conducted a Monte Carlo Markov Chain (MCMC) analysis employing the ECLAIR code~\citep{ilicDarkMatterProperties2021,foreman-mackeyEmceeMCMCHammer2013} with its default parametrisation (including the so-called "minimal" neutrino scenario~\footnote{That is to say only one massive neutrino of mass $0.06\,$eV.}) to fit our model to the latest Planck PR4 data~\citep{tristramCosmologicalParametersDerived2024}, including the latest lensing signal from~\cite{carronCMBLensingPlanck2022a}. More precisely, we used the likelihoods Hillipop TTTEEE, Lollipop low-$\ell$ E, and Commander for low TT (from P18) in combination with data from the BAO from DESI~\citep{desicollaborationDESI2024VI2024a}. In some places in the paper, we also use the latest SDSS consensus in order to compare with DESI. It is composed of the main galaxy sample (MGS) from DR7~\citep{rossClusteringSDSSDR72015a}, luminous red galaxies (LRG) from DR12~\citep{alamClusteringGalaxiesCompleted2017a}, and the DR16 release (emission line galaxies, ELG~\citep{raichoorCompletedSDSSIVExtended2020}, LRG~\citep{gil-marinCompletedSDSSIVExtended2020}, Lyman-$\alpha$ forest measurements, LYA~\citep{bourbouxCompletedSDSSIVExtended2020}, and quasars, QSO~\citep{houCompletedSDSSIVExtended2020}), in addition with data from the 6dF survey~\citep{beutler6dFGalaxySurvey2011a}. 

The measurements of $\lambda_G$ using different datasets and models are summarized in Table~\ref{tab:lambdaG}. For the baseline $\Lambda$CDM model, one obtains 
\begin{equation}
    G = \lambda_G^2 G_{\text{codata}} = (6.75\pm0.12)\times 10^{-11} \mathrm{m}^3\cdot\mathrm{kg}^{-1}\cdot\mathrm{s}^{-2}\;,
\end{equation}

\begin{figure*}[h!t]
    \centering
    \includegraphics{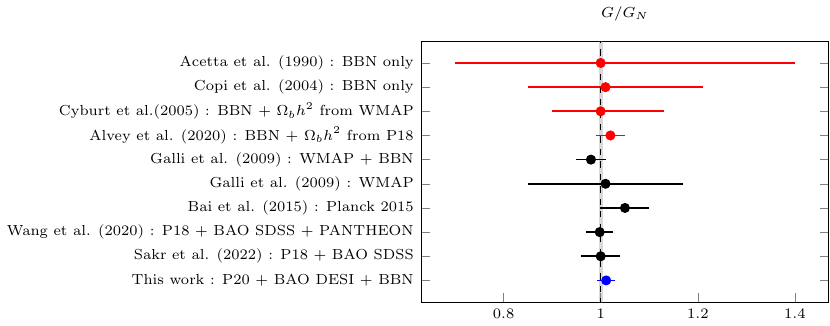}
    \caption{Recent constraints on cosmological measurements of $G/G_{\text{codata}}$ using high-redshift probes. The dashed vertical line corresponds to $G=G_{\text{codata}}$, while the light gray area corresponds to an ultimate cosmic-variance-limited precision that could be obtained with the CMB according to~\cite{PhysRevD.80.023508}.}
    \label{fig:whisker_G}
\end{figure*}

\begin{table*}[h!]
    \centering
    \caption{Constraints on a cosmological measurement of $G=\lambda_G^2G_{\text{codata}}$ for
      different data combination and cosmological models. The first two lines are obtained by fixing $\lambda_G=1$ in our code, and give consistent results with the literature.}
    \label{tab:lambdaG}
    \begin{tabularx}{\textwidth}{@{}rlccccc@{}}
        \hline
        model & data & $\lambda_G$ & $H_0$ (km.s$^{-1}$.Mpc$^{-1})$ & $w_0$ & $w_a$ & $\Omega_{\text{k}}$ \\\hline
        $\Lambda$CDM & P20 TTTEEE + BAO SDSS& $-$ & $67.74\pm0.41$ & $-$ & $-$ & $-$ \\
        $\Lambda$CDM & P20 TTTEEE + BAO DESI& $-$ & $68.22\pm0.39$ & $-$ & $-$ & $-$ \\
        $\lambda_G\Lambda$CDM & P20 TT & $1.048^{+0.027}_{-0.035}$ & $74.4^{+3.7}_{-5.2}$ & $-$ & $-$ & $-$\\
        $\lambda_G\Lambda$CDM & P18 TT + low E & $0.998\pm0.018$ & $66.9\pm2.00$ & $-$ & $-$ & $-$ \\
        $\lambda_G\Lambda$CDM & P18 TT + low E high-$\ell$ & $0.969^{+0.052}_{-0.058}$ & $62.0^{+3.9}_{-4.9}$ & $-$ & $-$ & $-$ \\
        $\lambda_G\Lambda$CDM & P18 TT + low E low-$\ell$ & $1.18^{+0.13}_{-0.038}$ & $80.9^{+8.5}_{-4.4}$ & $-$ & $-$ & $-$ \\
        $\lambda_G\Lambda$CDM & P20 EE & $1.062^{+0.052}_{-0.059}$ & $74.7^{+5.4}_{-6.2}$ & $-$ & $-$ & $-$ \\
        $\lambda_G\Lambda$CDM & P20 TE & $1.023\pm0.047$ & $70.5\pm4.4$ & $-$ & $-$ & $-$ \\
        $\lambda_G\Lambda$CDM & P20 TTTEEE & $1.007\pm0.011$ & $68.3\pm1.1$ & $-$ & $-$ & $-$ \\
        $\lambda_G\Lambda$CDM & P20 TTTEEE + BAO DESI& $1.010\pm0.011$ & $69.11\pm0.98$ & $-$ & $-$ & $-$ \\
        $\lambda_G\Lambda$CDM & P20 TTTEEE + BAO DESI + BBN & $1.0056\pm0.0091$ & $68.63\pm0.87$ & $-$ & $-$ & $-$ \\
        $\lambda_Gw_0$CDM & P20 TTTEEE + BAO DESI& $1.008\pm0.011$ & $70.8^{+1.7}_{-1.9}$ & $-1.070^{+0.062}_{-0.055}$ & $-$ & $-$ \\
        $\lambda_Gw_0w_a$CDM & P20 TTTEEE + BAO DESI& $1.006\pm 0.011$ & $64.9 \pm 3.5$ & $-0.46^{+0.31}_{-0.42}$ & $-1.67^{+1.2}_{-0.79} $ & $-$ \\
        $\lambda_G\Omega_{\text{k}}\Lambda$CDM & P20 TTTEEE + BAO DESI& $1.008\pm0.011$ & $69.3\pm1.0$ & $-$ & $-$ & $0.0016\pm0.0015$ \\
        \hline
    \end{tabularx}
\end{table*}

\noindent which corresponds to a $1.8\%$ measurement of $G$, in accordance with the CODATA value~\citep{codata} within $1\sigma$ confidence interval. Notably, the precision is more than twice better than the one achieved with Planck 2015~\citep{Bai2015}, and nearly ten times more precise than that obtained with WMAP~\citep{PhysRevD.80.023508}. It is also $\sim25\%$ better than the precision obtained with Planck 2018 + BAO~\citep[see for example][]{wangConstraintsNewtonConstant2020b}. However, this level of precision still falls short of the ultimate, variance-limited measurement of $G$ via the CMB, projected to be $0.4\%$ according to forecasts presented in~\cite{PhysRevD.80.023508}. This suggests the potential for enhancement through forthcoming generations of CMB missions~\citep{ballardiniCosmologicalConstraintsGravitational2022a}. A summary of recent cosmological constraints on $G$ using high-redshift probes is provided in the whisker plot of Fig.~\ref{fig:whisker_G}.

It is worth noting that neglecting the PArthENoPE correction for the BBN would not change much the conclusion, and simply increase the uncertainty in the determination of $G$. As an illustrative example, for P20 only, one would obtain $\lambda_G=1.010 \pm 0.013$ without the BBN correction, instead of $\lambda_G=1.007 \pm 0.011$ with the BBN correction taken into account (see Table~\ref{tab:lambdaG}).

\begin{figure}[h!t]
    \centering
    \includegraphics[width=\linewidth]{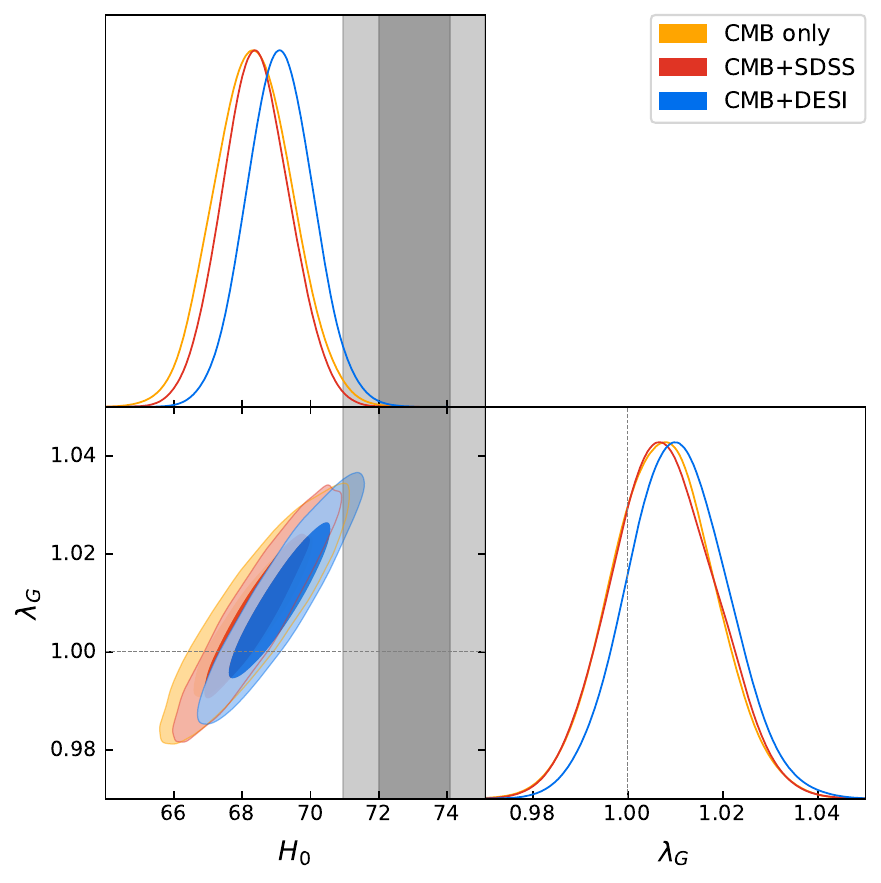}
    \caption{Constraints and correlation between $\lambda_G$ and $H_0$ with CMB+BAO data. The constraint on $\lambda_G$ is only marginally improved when adding the BAO. The DESI BAO slightly shift the central value of $H_0$ towards higher values, leading to a moderate $2.8\sigma$ tension with the local measurements of $H_0$ indicated by the vertical grey line~\citep{riessComprehensiveMeasurementLocal2022b}.}
    \label{fig:lambda_G_CMB_BAO}
\end{figure}

Upon examination of Table~\ref{tab:lambdaG} and Fig.~\ref{fig:lambda_G_CMB_BAO}, we observe that the BAO data have
a modest impact on the constraint for $\lambda_G$. They marginally 
nudge $\lambda_G$ closer to unity while slightly reducing its uncertainty. Figure~\ref{fig:lambda_G_CMB_BAO} also reveals a correlation between $\lambda_G$ and the Hubble constant $H_0$, highlighting a connection between the gravitational constant and the Hubble constant. Nevertheless, as it can be seen from Fig.~\ref{fig:lambda_G_CMB_BAO}, a modified value of $G$ at cosmological scales cannot fully resolve the Hubble tension~\citep{Verde_2019}, even tough the tension decreases to $2.8\sigma$ due to a slight increase of the central value of $H_0$ as well as an increase by a factor $\sim2$ of the uncertainty. This result is consistent with previous studies in the literature~\citep{wangConstraintsNewtonConstant2020b,sakrCanVaryingGravitational2022}.

In Table~\ref{all_constraints}, we present the constraints on the main cosmological parameters derived from an analysis using the Planck 2020 dataset only. Notably, the inclusion of the free parameter $\lambda_G$ primarily contributes to an increase of the uncertainty associated with each parameter when compared to the baseline $\Lambda$CDM model.
\begin{table}[h!t]
    \begin{center}
        \caption{\label{all_constraints} Constraints on the standard cosmological parameters using Planck PR4 (denoted as P20) for the $\lambda_G\Lambda$CDM and $\Lambda$CDM models. The uncertainty on all cosmological parameters (except $A_{\text{s}}$ and $\tau_{\text{reio}}$) are increased with the addition of the free parameter $\lambda_G$.}
        \begin{tabular}{ccc}
            \hline
             Parameter & $\lambda_G\Lambda$CDM P20 & $\Lambda$CDM P20\\\hline
            $\omega_{\text{b}}$ & $0.02264 \pm 0.00052$ & $0.02224 \pm 0.00013$\\
            $\omega_{\text{cdm}}$ & $0.1193 \pm 0.0024$ & $0.1188 \pm 0.0012$\\
            $H_0$ & $68.3 \pm 1.1$ & $67.66 \pm 0.52$\\
            $\tau_{\text{reio}}$ & $0.0582 \pm 0.0063$ & $0.0580 \pm 0.0062$\\
            $\ln(10^{10}A_{\text{s}})$ & $3.042 \pm 0.015$ & $3.040 \pm 0.014$\\
            $n_{\text{s}}$ & $0.9709 \pm 0.0063$ & $0.9680 \pm 0.0040$\\
            \hline
        \end{tabular}
    \end{center}
\end{table}

\begin{figure}[ht]
  \centering
  \includegraphics[width=\linewidth]{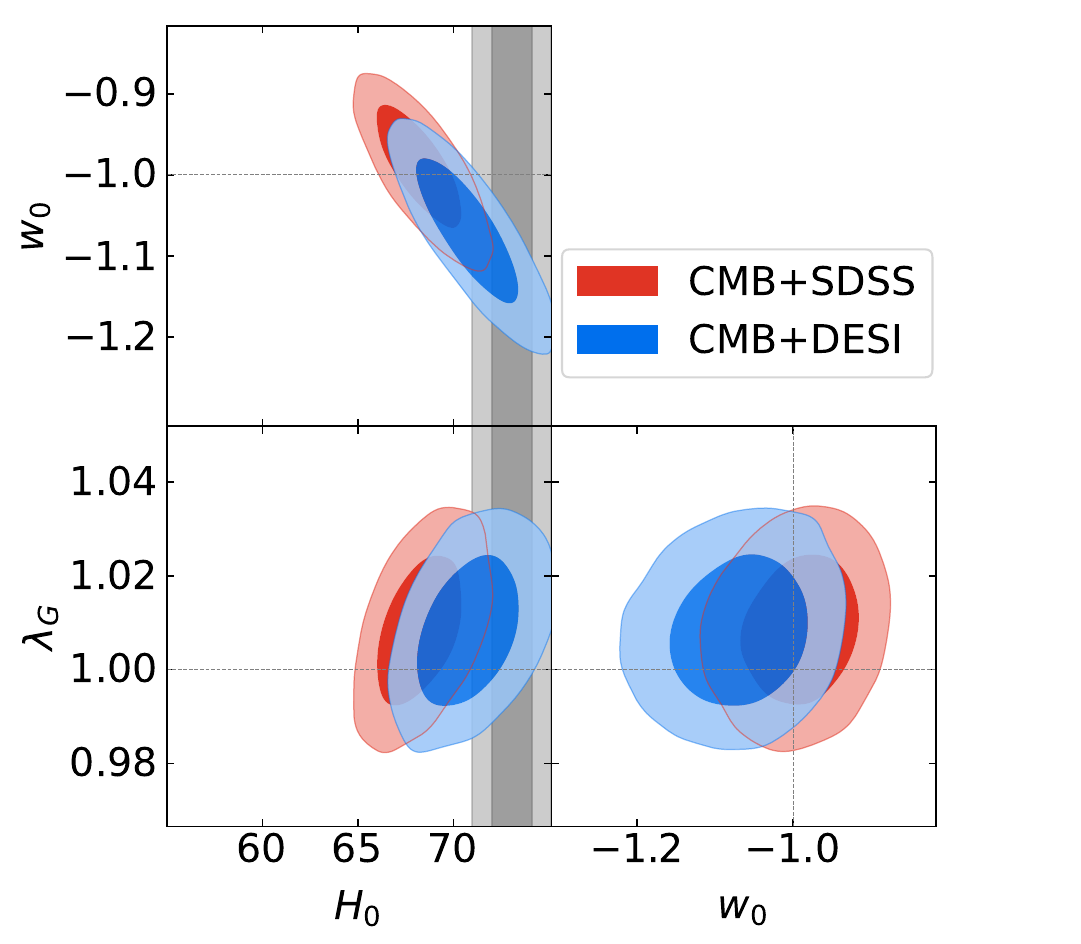}
  \caption{Constraints on $\lambda_G$ and $w_0$ for the $\lambda_Gw_0$CDM model. The measurement of $H_0$ from~\cite{riessComprehensiveMeasurementLocal2022b} is represented as a grey vertical band.}\label{fig:wCDM}
\end{figure}

\begin{figure}[ht]
  \centering
  \includegraphics[width=\linewidth]{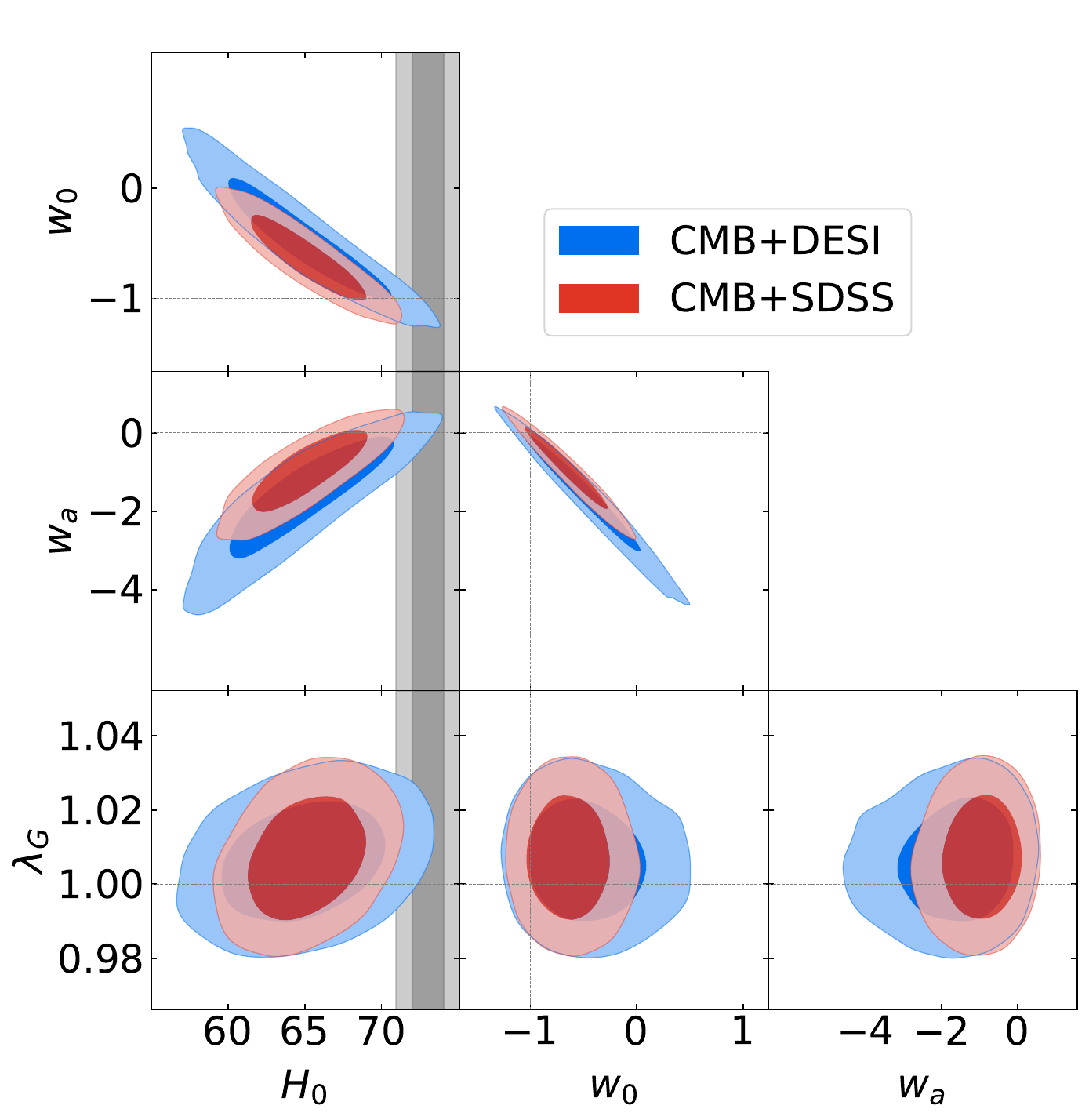}
  \caption{Constraints on $(\lambda_G,w_0,w_a)$ for the $\lambda_Gw_0w_a$CDM model. The measurement of $H_0$ from~\cite{riessComprehensiveMeasurementLocal2022b} is represented as a grey vertical band.}\label{fig:w0waCDM}
\end{figure}

\begin{figure}[ht]
  \centering
  \includegraphics[width=\linewidth]{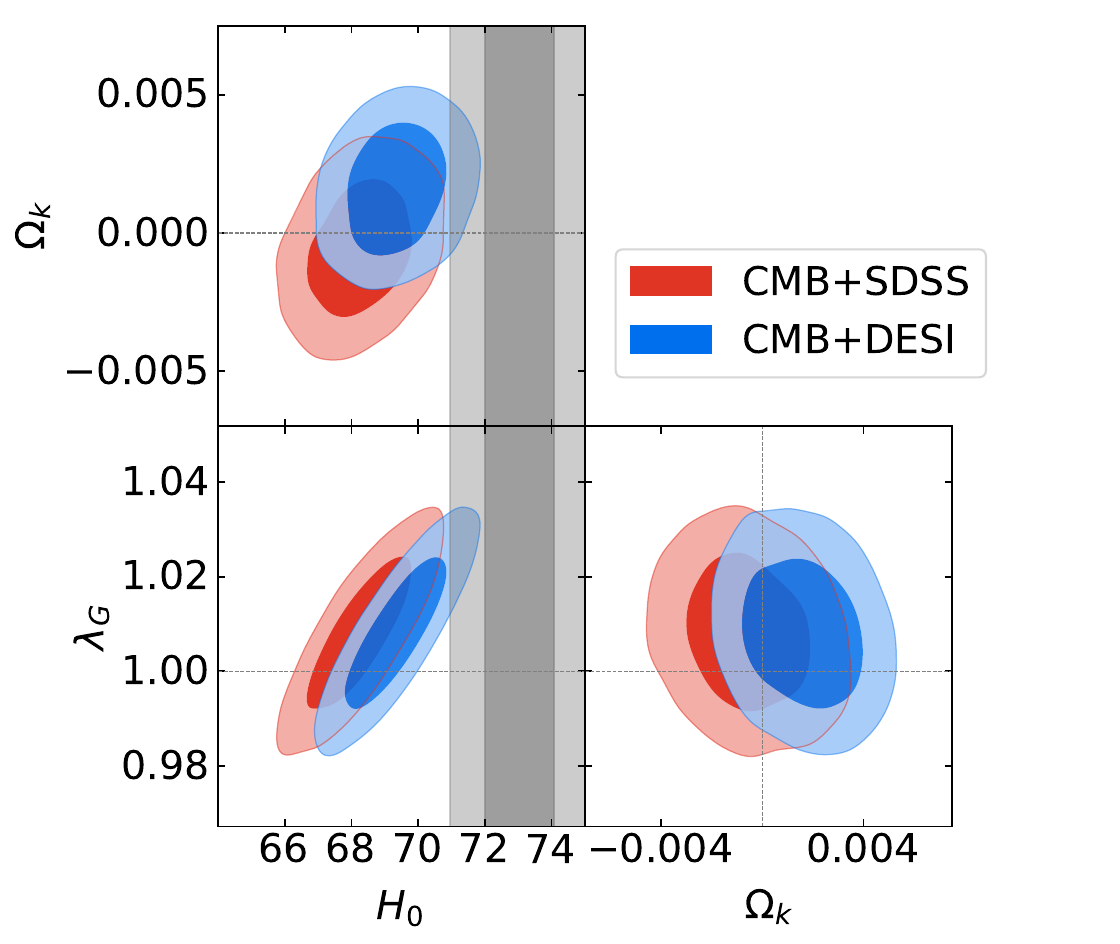}
  \caption{Constraints on $\lambda_G$ and $\Omega_{\text{k}}$ for the $\lambda_G\Omega_{\text{k}}\Lambda$CDM model. The measurement of $H_0$ from~\cite{riessComprehensiveMeasurementLocal2022b} is represented as a grey vertical band.\label{fig:OmegakCDM}}
\end{figure}

In order to assess the dependence of the cosmological measurement of $G$ on the choice of the cosmological model, we conducted the analysis within various extensions of the $\Lambda$CDM framework. Specifically, we first explored the $w_0$CDM model, which accommodates a more general description of the dark energy fluid by introducing a free equation of state parameter, $w_0=p/\rho$. In the context of this $\lambda_Gw_0$CDM model, a notable correlation emerges between $w_0$ and $H_0$, leading to an increase in the uncertainty associated to $H_0$ (see Fig.~\ref{fig:wCDM}, where we also show the difference between the BAO from the SDSS consensus and the BAO from DESI). We also considered the more sophisticated CPL dark energy equation of state $w(z)=w_0+w_a\frac{z}{1+z}$~\citep{chevallierAcceleratingUniversesScaling2001,linderExploringExpansionHistory2003}, named $\lambda_Gw_0w_a$CDM model. As it can be seen in Table~\ref{tab:lambdaG}, the constraint on $\lambda_G$ is not affected by this generalized dark energy equation of state. One also recovers the result that an evolving dark energy is marginally favored~\citep{desicollaborationDESI2024VI2024a}, even when allowing $G$ to vary, as it can be seen from Fig.~\ref{fig:w0waCDM}. Finally, we considered spatially curved models, where $\Omega_{\text{k}}$ serves as a free parameter. For this non-flat model, the inclusion of BAO data is imperative to mitigate significant parameter degeneracies. The outcomes of this study are summarized in the final four rows of Table~\ref{tab:lambdaG} and indicates that $G$ remains largely unaffected by the assumptions on the cosmological model. As illustrated in Figs.~\ref{fig:wCDM},~\ref{fig:w0waCDM}, and~\ref{fig:OmegakCDM}, these additional parameters are consistent with the standard model, indicating no curvature ($\Omega_{\text{k}}$ compatible with zero within $1\sigma$, inline with the latest P20 result~\cite{tristramCosmologicalParametersDerived2024}) and a pure cosmological constant ($(w_0,w_a)$ consistent with $(-1,0)$ within $1\sigma$).

\begin{figure}[h!t]
    \centering
    \includegraphics[width=\linewidth]{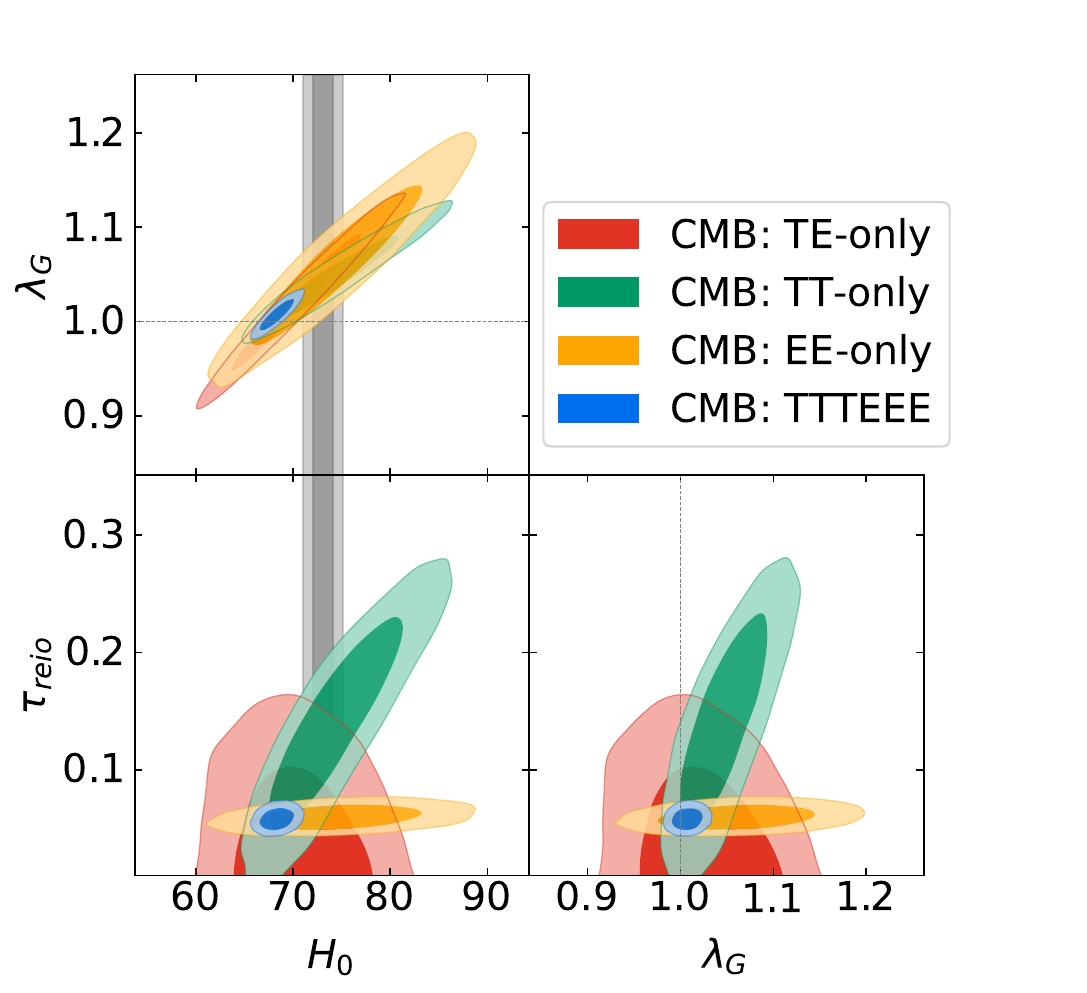}
    \caption{Influence of different CMB spectra in the constraints and correlation between $\lambda_G$ and $H_0$. The measurement of $H_0$ from~\cite{riessComprehensiveMeasurementLocal2022b} is represented as a grey vertical band.}
    \label{fig:lambda_G_CMB}
\end{figure}

This section concludes with a detailed examination of the distinct constraints derived from various subsets of the CMB data. Notably, as already mentioned earlier, the primary source of constraints on $\lambda_G$ stems from its interplay with polarization data. Specifically, when considering the temperature spectrum (TT) alone or the polarization spectrum (EE) alone, there is a preference for a larger value of $G$, which subsequently yields a higher estimate for the Hubble constant $H_0$. This preference aligns with the local measurements reported in~\cite{riessComprehensiveMeasurementLocal2022b}, as depicted in Fig.~\ref{fig:lambda_G_CMB} and detailed in Table~\ref{tab:lambdaG}. However, those preferences are accompanied by broader uncertainty ranges. The TE spectrum, on the other hand, occupies an intermediate position between the high and low value for $H_0$ (see Fig.~\ref{fig:lambda_G_CMB} and Table~\ref{tab:lambdaG}). Notably, the preferred value of $H_0$ with EE alone changed significantly from P18 to P20. Specifically, using P18 data results in $H_0=65.1^{+6.8}_{-8.8}\,\text{km}\cdot\text{s}^{-1}\cdot\text{Mpc}^{-1}$ compared to $74.7^{+5.4}_{-6.2}\,\text{km}\cdot\text{s}^{-1}\cdot\text{Mpc}^{-1}$ of P20 (see Table~\ref{tab:lambdaG}). The same is true for the central value of $\lambda_G$, going from $0.956^{+0.069}_{-0.087}$ in P18 to $1.062^{+0.052}_{-0.059}$ in P20. The reason why the combination of TT and EE can give a smaller value of $H_0$ is illustrated in Fig.~\ref{fig:lambda_G_CMB}, where the correlation between $\lambda_G$ and $H_0$ and the optical depth of reionization $\tau_{\text{reio}}$, is evident. The intersection of the contours in the $(\tau_{\text{reio}},H_0)$ plot for TT and EE explains the low value of $H_0$. The better determination of this parameter $\tau_{\text{reio}}$ in the Planck's PR4 analysis enhances the constraining power of combining TT and EE data. A further improved determination of $\tau_{\text{reio}}$ is therefore promising for a more precise measurement of $G$ at cosmological scale (see for example~\cite{montero-camachoFiveParametersAre2024} who propose a method to improve the determination of the optical depth of reionization from the cosmology itself). As a conclusion of this paragraph, it is only through the combination of all these distinct CMB data probes that we can achieve a precise measurement of $G$ and obtain a Hubble constant in accordance with the Planck 2020 dataset.

\begin{figure}[h!t]
    \centering
    \includegraphics[width=\linewidth]{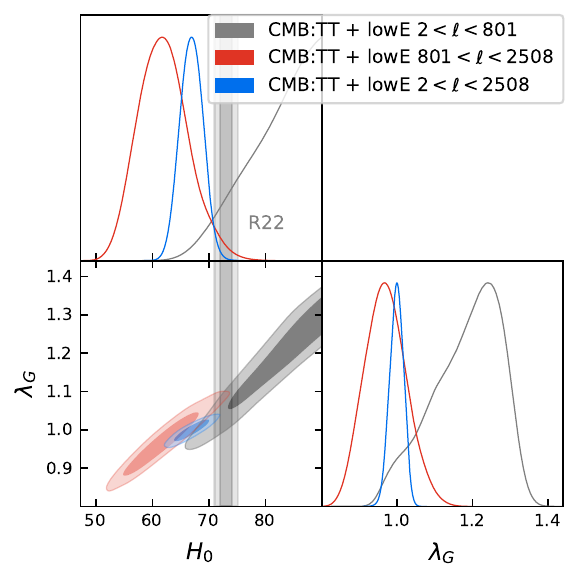}
    \caption{High-$\ell$ versus low-$\ell$ constraints and correlation between $\lambda_G$ and $H_0$ when using only the temperature spectrum of Planck 2018. Note that the analysis is done with Planck 2018, for which the likelihoods divided in $\ell$ are available. The vertical grey band correspond to the Riess value~\citep{riessComprehensiveMeasurementLocal2022b}.}
    \label{fig:lambda_G_CMB_ell}
\end{figure}

One of the reasons behind the poor constraint obtained from the TT spectrum lies in the presence of degeneracies with other cosmological parameters that can mimic the same physical effect as a modification of $G$, namely altering the damping for large $\ell$. It turns out that a combination with only the lowE part of the polarization spectrum data can already help break these degeneracies. In more details, the combination TT+lowE data greatly enhances the determination of cosmological parameters, and in particular a factor of two on $\lambda_G$, as demonstrated in Table~\ref{tab:lambdaG} and Fig.~\ref{fig:lambda_G_CMB_ell} (where the blue curve represents the TT+lowE part of the Planck 2018 data).

Within the TT+lowE subset of the CMB data, it is instructive to distinguish between the contributions from low-$\ell$ ($\ell<\ell^*$) and the high-$\ell$ part ($\ell>\ell^*$) since the effect of $\lambda_G$ is different on the small and high scales, as discussed in Sect.~\ref{recombination}. Notably, the high-$\ell$ values exhibit more pronounced damping with increasing $G$ and the low-$\ell$ part is relatively insensitive to variations in $G$ (or can at least be absorbed by adjusting the initial amplitude of the matter power spectrum, as discussed in Sect.~\ref{recombination}). It is therefore unsurprising that the constraint on $\lambda_G$ is significantly degraded when considering the low-$\ell$ data only. This low-$\ell$ part tends to favor a higher value of $\lambda_G$ and consequently this also results in a higher determination of $H_0$, as depicted in Fig.~\ref{fig:lambda_G_CMB_ell}. On the other hand, when focusing on the high-$\ell$ part of the spectrum, a similar outcome is observed, but this time with a preference for a smaller value of $\lambda_G$ and a lower estimate for $H_0$. Note finally that even within the $\Lambda$CDM model, there is a well-known discrepancy between the low-$\ell$ and high-$\ell$ portions in terms of the derived value for $H_0$. However, this discrepancy is typically considered non-significant enough to be deemed problematic~\citep{planckcollaborationPlanckIntermediateResults2017}. Nevertheless, within the $\lambda_G$ model, this inconsistency becomes even larger and should be investigated within the latest PR4 data release.

\section{Conclusion}\label{conclusions}

In this paper, we demonstrate that the impressive precision of cosmic microwave background (CMB) observations allows us to measure the gravitational constant $G$ with a remarkable $2.1\%$ degree of precision. In turns out that this measurement does not depend on model assumptions such as the nature of dark energy or the presence of spatial curvature. Moreover, it is independent from laboratory
experiments and probes $G$ on cosmological scales that extend far beyond the reach of laboratory measurements. Our findings yield a value of $G = 6.82\pm0.14 \times 10^{-11} \text{m}^3\cdot\text{kg}^{-1}\cdot\text{s}^{-2}$, which aligns seamlessly with laboratory measurements and corresponds to a $\sim25\%$ improvement compared to the literature. Importantly, our study suggests that forthcoming missions, which aim to enhance polarization measurements~\citep{litebirdcollaborationProbingCosmicInflation2023,abazajianCMBS4ScienceCase2019}, hold the potential to further refine and bolster this cosmological-scale measurement of $G$. Finally, we conclude this paper by noting that the constraint obtained on the measurement of the gravitational constant is valid only if $G$ is the sole fundamental constant that varies. The situation differs if, for instance, both $G$ and the fine structure constant $\alpha$ are allowed to vary simultaneously.

\begin{acknowledgements}
We thank James Rich for fruitful discussions on the revised version of the paper. We acknowledge use of CLASS (\url{https://github.com/lesgourg/class_public}) for calculating power spectra and ECLAIR (\url{https://github.com/s-ilic/ECLAIR}) for the sampling of the likelihoods.
\end{acknowledgements}

%
   \bibliographystyle{aa} 
   \bibliography{Biblio} 
%

\end{document}